\documentclass[12pt]{article}
\usepackage{jheppub}
\usepackage{amsmath, amssymb, mathtools}
\usepackage{enumitem}

\setenumerate{label=\arabic*)}

\usepackage{ytableau}
\ytableausetup{boxsize=.6em,centertableaux}
\newcommand{\anti}{\ydiagram{1,1}}
\newcommand{\sym}{\ydiagram{2}}

\def\be#1\ee{\begin{align}#1\end{align}}
\def\eqn{\eqref}
\def\nn{\nonumber}

\newcommand{\al}{\alpha}
\newcommand{\te}{\theta}
\newcommand{\la}{\lambda}

\newcommand{\onov}[1]{\frac{1}{#1}}
\newcommand{\mat}[1]{\left(\begin{matrix}#1\end{matrix}\right)}

\newcommand{\lag}{\mathcal{L}}

\def\Z{\mathbb{Z}}

\def\vev#1{\langle #1 \rangle}

\def\deltafix{\kern-.1em}

\title{On Discrete Anomalies in Chiral Gauge Theories}

\author{Philip Boyle Smith, Avner Karasik, Nakarin Lohitsiri and David Tong}

\affiliation{Department of Applied Mathematics and Theoretical Physics \\ University Cambridge, CB3 0WA, UK}

\emailAdd{pb594@cam.ac.uk, avnerkar@gmail.com, nl313@cam.ac.uk, d.tong@damtp.cam.ac.uk}

\abstract{We study two well-known $SU(N)$ chiral gauge theories with fermions in the symmetric, anti-symmetric and fundamental representations. We give a detailed description of the global symmetry, including various discrete quotients. Recent work argues that these theories exhibit a subtle mod 2 anomaly, ruling out certain phases in which the theories confine without breaking their global symmetry, leaving a gapless composite fermion in the infra-red. We point out that no such anomaly exists. We further exhibit an explicit path to the gapless fermion phase, showing that there is no kinematic obstruction to realising these phases.}

\begin{document}

\maketitle

\section{Introduction}

There is much that we don't know about chiral gauge theories. The Nielsen-Ninomiya theorem \cite{nn} means that there is no Monte Carlo safety net, making it challenging to get a handle on the dynamics of these theories at strong coupling. In this short note we comment on two of the simplest, and best studied, chiral gauge theories which we refer to as the ``symmetric'' and ``anti-symmetric'' theories respectively.

\vspace{1em}
\noindent
\underline{The Symmetric Theory}

\noindent
The first is an $SU(N)$ gauge theory with left-handed Weyl fermions in the following representations:
\be \mbox{Symmetric $\sym$ and $(N+4)$ anti-fundamentals $\overline{\Box}$} \label{sym}\ee
We refer to the $\sym$ fermion as $\lambda$ and the $\overline{\Box}$ as $\psi$. Ignoring subtleties such as discrete quotients (which will be important later) the global symmetry is
\be G = SU(N+4) \times U(1) \label{symsym}\ee
The theory flows to strong coupling in the infra-red and there are a number of different proposals for the low-energy physics
\begin{itemize}
\item The theory may confine without breaking the global symmetry group $G$. In this case, there must be massless fermions to saturate the 't Hooft anomalies of $G$. The most obvious candidate is the composite $\lambda\psi\psi$, transforming in the ${\anti}$ of $SU(N+4)$ \cite{raby}. In what follows, we will refer to this as the ``confining phase''.
\item The global symmetry group may be broken to $G\rightarrow SU(N) \times SU(4) \times U(1)'$. This occurs, for example, if there is a gauge non-invariant condensate $\langle \lambda \psi \rangle \neq 0$. In what follows, we will refer to this, somewhat inaccurately, as the ``Higgs phase''.
\item More recently, a study of softly broken supersymmetric theories suggested different behaviour. For $N \geq 13$, it was proposed that the symmetry group is broken to $G \rightarrow SO(N+4)$ while, for $N < 13$ the theory was argued to flow to an interacting infra-red fixed point \cite{mura2}.
\end{itemize}

\vspace{1em}
\noindent
\underline{The Anti-Symmetric Theory}

\noindent
The second theory again has $SU(N)$ gauge group, this time with left-handed Weyl fermions transforming as
\be \mbox{Anti-symmetric $\anti$ and $(N-4)$ anti-fundamentals $\overline{\Box}$} \label{anti}\ee
We refer to the $\anti$ fermion as $\chi$ and the $\overline{\Box}$ as $\psi$. Ignoring discrete subtleties, the global symmetry group is now
\be G = SU(N-4) \times U(1) \label{symanti}\ee
There are, again, different proposals for the infra-red physics.
\begin{itemize}
\item This theory may again confine without breaking the global symmetry $G$. The perturbative 't Hooft anomalies are saturated by the massless composite fermion $\chi\psi\psi$, transforming in the ${\sym}$ representation of $SU(N-4)$.

This time, something different happens in the would-be Higgs phase. If the gauge non-invariant bilinear $\langle \chi\psi\rangle$ is assumed to get an expectation value, then the global symmetry $G$ remains unbroken, albeit after twisting with the $SU(N)$ gauge symmetry. Once again, one finds a single massless fermion, transforming in the ${\sym}$ of the $SU(N-4)$ global symmetry. This means that the putative confining and Higgs phases are thought to yield the same low-energy physics in this theory, an observation that was referred to as {\it complementarity} \cite{raby}.

\item The study of softly-broken supersymmetric theories again suggests different low-energy dynamics. For $N$ even, the proposal is that $SU(N-4) \rightarrow Sp(\frac{1}{2}(N-4))$ (using the convention $Sp(1) \equiv SU(2)$) with only Goldstone bosons in the low-energy spectrum. For $N$ odd, the suggestion is that $SU(N-4) \rightarrow Sp(\frac{1}{2}(N-5))$, this time with both Goldstone bosons and massless fermions to saturate the 't Hooft anomalies \cite{mura1}.
\end{itemize}

\noindent
All the proposals above for both the symmetric and anti-symmetric theories are conjectures. It may be that the true dynamics of these theories is something different yet again. Further discussion of the phases of these theories can be found in \cite{san}.

\subsubsection*{Can Discrete Anomalies Help?}

The array of different options for these theories reflects our lack of control over the low-energy physics. Clearly, it would be extremely useful if there were some way to get a better handle on the strong coupling regime.

Outside of supersymmetric theories, one of the few handles that we have at strong coupling comes from symmetries and their attendant 't Hooft anomalies. All phases described above are consistent with perturbative 't Hooft anomalies, but in recent years it has been appreciated that precious information about 4d gauge theories can be harvested from anomalies associated to discrete symmetries or to generalised higher-form symmetries \cite{seiberg2,qcd4}. It is natural to ask: can these techniques be brought to bear on the chiral theories above?

This question was answered in the affirmative in a pair of papers \cite{bk2,bk1}. The authors performed a detailed study of the global symmetry group, dressing \eqn{symsym} and \eqn{symanti} with a number of discrete quotients. They then argued for a mixed anomaly between $(-1)^F$ and a $\mathbb{Z}_N$ 1-form symmetry that, for $N$ even, could be used to reach the following conclusions:
\begin{enumerate}
\item The confining phase of the symmetric theory, with unbroken $G=SU(N+4) \times U(1)$ is inconsistent with discrete anomaly matching.
\item The confining phase of the anti-symmetric theory is distinct from the Higgs phase, even though both have global symmetry group $G=SU(N-4) \times U(1)$ and the same massless matter content. Moreover, the confining phase is inconsistent with discrete anomaly matching while the Higgs phase is not.
\end{enumerate}

If true, these would be important results. However there are reasons to be cautious. In particular, the claim that the Higgs and confining phases of the anti-symmetric theory are distinct is striking given that these theories have the same gapless spectrum and the same symmetry including, as we will review below, the discrete quotient structure.

In this short note, we revisit the discrete anomaly calculation of \cite{bk1,bk2}. We show that the claimed mod 2 anomaly is an illusion. It does not exist. Correspondingly, there is evidence for neither claim 1) nor 2). The confining phase of the theories cannot be ruled out, at least not using methods wielded to date.

Sadly therefore, our paper serves primarily to increase, rather than reduce, our ignorance about these theories. We will, however, describe in some detail the structure of the discrete symmetries and their consequences. This includes both physical manifestations of the symmetry structure, such as the spectrum of line and surface operators, and more mathematical renderings of the subject including a computation of the relevant cobordism group. In all approaches, we see that there is no discrete anomaly at play. Furthermore, in Section \ref{otherargs}, we describe a deformation of each theory that results in the spectrum of the confining phase, showing explicitly that there is no obstacle to realising these phases at low energies.

\section{Aspects of Discrete Symmetry} \label{putative}

In this section, we give a careful analysis of the symmetry group and possible discrete anomalies associated to the chiral theories. For concreteness, we will focus on the anti-symmetric theory. However, all arguments below are easily adapted for the symmetric theory.

\subsection{The Symmetry Structure}

As we explained in the introduction, the anti-symmetric theory consists of an $SU(N)$ gauge group, with a Weyl fermion $\chi$ in the $\anti$ representation and $N-4$ fermions $\psi$ in the $\overline{\Box}$ representation. The naive global symmetry group is
\be G= SU(N-4)\times U(1)\label{naive}\ee
with the $SU(N-4)$ rotating the $\psi$ fermions. However, a more careful study involves a number of discrete quotients \cite{bk1,bk2}, as we now review.

The charges under the global non-anomalous $U(1)$ depend on whether $N$ is odd or even. For $N$ odd, we have $q_\chi=N-4$ and $q_\psi=2-N$. For $N$ even, we must divide by 2 to ensure that the $U(1)$ acts faithfully on the matter content. Because the putative anomaly described in \cite{bk1,bk2} arises only for $N$ even, we restrict to this case and take
\be q_\chi = \frac{N-4}{2} \ \ \ {\rm and} \ \ \ q_\psi = \frac{2-N}{2} \nn\ee
The naive symmetry group \eqn{naive} correctly captures the Lie-algebraic part of the global symmetry but has two shortcomings. First, there could be additional disconnected components. Under a general $U(1)_\chi\times U(1)_\psi$ transformation, the theta parameter is shifted by
\be \chi\to e^{2\pi i \omega_\chi}\chi\ ,\ \psi\to e^{2\pi i\omega_\psi}\psi\ \Rightarrow\ \delta\te=2\pi (N-2)\omega_\chi+2\pi (N-4)\omega_\psi \ee
In addition to the global $U(1)$ under which $\delta\theta=0$, there might be discrete transformations that change $\theta$ by $2\pi\mathbb{Z}$. Taking $\omega_\chi=-\omega_\psi=\onov{2}$ we get
\be \chi,\psi\to -\chi,-\psi\Rightarrow\delta\te=2\pi\nn\ee
This is the only additional component of the symmetry group and it clearly coincides with fermion parity, $(\mathbb{Z}_2)_F=(-1)^F$.

The second shortcoming is that some elements of the symmetry group may act trivially on fermions, or coincide with a gauge transformation. To find these redundant elements, consider the most general centre transformation of $SU(N) \times SU(N-4)$, accompanied by a $U(1)$ transformation. The effect on the fermions is
\be \psi \ &\to \ e^{-2\pi im/N} e^{2\pi i k/(N-4)} e^{\pi i (2-N) \al} \psi \nn\\ \chi \ &\to \ e^{4\pi im/N}e^{\pi i(N-4)\al} \chi \ \label{center}\ee
where $m, k \in \mathbb{Z}$ parametrise the centre of $SU(N)$ and $SU(N-4)$ respectively and $\alpha$ parametrises the $U(1)$ action. For even $N$, it is not possible to eliminate a general centre transformation by a $U(1)$ rotation. However, for any centre transformation it is always possible to find a $U(1)$ rotation such that the combination acts as fermion parity $(-1)^F$. For the $SU(N)$ centre transformation $(m = 1, k = 0)$, we can take the $U(1)$ transformation to be
\be \al=\frac{1}{N}\ \ \ \Longrightarrow\ \ \ \psi,\chi \rightarrow -\psi,-\chi\label{alpha1}\ee
For an $SU(N-4)$ center transformation $(m = 0, k = 1)$ we can take
\be \al=\onov{N-4}\ \ \ \Longrightarrow\ \ \ \psi,\chi\rightarrow -\psi,-\chi \label{alpha2}\ee
We should then quotient by these transformations to obtain a faithfully acting symmetry group. The correct refined version of \eqn{symanti} is, for $N$ even \cite{bk1,bk2},
\be G = \frac{SU(N-4) \times U(1) \times (\mathbb{Z}_2)_F}{\mathbb{Z}_N \times \mathbb{Z}_{N-4}} \quad\quad N\ {\rm even}\label{geven}\ee
We will shortly couple background gauge fields to this symmetry group. However, before we proceed it will be useful to make a number of further remarks about the presence of the $(\mathbb{Z}_2)_F$ in the symmetry group \eqn{geven}.

First, $(\mathbb{Z}_2)_F$ also sits within the $\text{Spin}(4)$ spacetime symmetry. This becomes important when the theory is placed on curved backgrounds and we will return to this point in Section \ref{boredsec}. For now, however, we can restrict our discussion to flat space.

Second, the discussion above does not take full advantage of the fact that the $SU(N)$ action is a gauge symmetry (read redundancy) in whittling down the global symmetry group to its simplest form. The first condition \eqn{alpha1} means that, up to a gauge redundancy, the $(\mathbb{Z}_2)_F$ becomes part of the $U(1)$. It should therefore be possible to give an alternative characterisation of $G$ without reference to $(\mathbb{Z}_2)_F$. As we now explain, the result is:
\be &G = \frac{SU(N - 4) \times U(1)}{\mathbb{Z}_{N(N-4)/4} \times \mathbb{Z}_2} \quad \quad \text{$N = 0$ mod 4}\label{gtrue}\\ &G = \frac{SU(N - 4) \times U(1)}{\mathbb{Z}_{N(N-4)/2}} \quad \quad \text{$N = 2$ mod 4} \nn\ee
To see this, note that from \eqn{center} the centre transformations that act trivially on the fermions $\chi,\ \psi$ are
\be \al=\frac{m}{N}+\frac{k}{N-4}\ \ \ {\rm with} \ \ \ m+k\in 2\mathbb{Z}\label{alpha3}\ee
This condition on $m+k$ implies that the quotient group must have dimension $\frac{1}{2}N(N-4)$. For $N=2$ mod $4$, the element $(m,k)=(1,1)$ is of order $\frac{1}{2}N(N-4)$, hence the quotient must be $\mathbb{Z}_{N(N-4)/2}$. For $N=0$ mod $4$, there is no element of order $\frac{1}{2}N(N-4)$. For example, the element $(m,k)=(1,1)$ is of order $\frac{1}{4}N(N-4)$ since $(m,k)$ are defined mod $(N,N-4)$. In addition we have the $\mathbb{Z}_2$ element $(m,k)=\left(\frac{N}{2},\frac{N-4}{2}\right)$ which is not a multiple of $(m,k)=(1,1)$. This is enough to deduce that the quotient group in this case is $\mathbb{Z}_{N(N-4)/4}\times \mathbb{Z}_2$.

\subsection{Adding Background Gauge Fields}

The existence of the discrete quotients in the global symmetry group brings a new opportunity. When the theory is coupled to background gauge fields for $G$, we can ask how the theory responds when the discrete gauge fields associated to the quotient take different values. This is how one detects more subtle discrete 't Hooft anomalies associated to the symmetry, as used to great effect in, for example, \cite{seiberg2,qcd4}.

The authors of \cite{bk1,bk2} apply this idea for the global symmetry group \eqn{geven} and reach a startling conclusion: they find two new anomalies, one between $(\mathbb{Z}_2)_F$ and $\mathbb{Z}_N$, and the other between $(\mathbb{Z}_2)_F$ and $\mathbb{Z}_{N-4}$. Furthermore, they claim that these anomalies cannot be matched by the conjectured massless fermions in the phase where the theory confines and $G$ is unbroken.

As explained above, the alternative form of the symmetry group \eqn{gtrue} does not exhibit a quotient by $\mathbb{Z}_N\times \mathbb{Z}_{N-4}$. Already, this casts some doubt on the claim that there is a mixed anomaly between $(\mathbb{Z}_2)_F$ and $\mathbb{Z}_N$ and between $(\mathbb{Z}_2)_F$ and $\mathbb{Z}_{N-4}$ as one can rephrase the symmetry without reference to these subgroups. Nonetheless, following \cite{bk1,bk2}, we return to the form of the symmetry group \eqn{geven} and introduce background gauge fields to see how the theory responds. To this end, we denote the original dynamical $SU(N)$ gauge field as $a$ and introduce background gauge fields $A$ for $U(1) \subset G$ and $A_f$ for $SU(N-4) \subset G$.

The two $SU(N)$ and $SU(N-4)$ gauge fields are then promoted to $U(N)$ and $U(N-4)$ gauge fields, denoted by $\tilde{a}$ and $\tilde{A}_f$ respectively, by coupling them to suitable higher form gauge fields: $B_c^{(1)}$ for $\mathbb{Z}_N$ and $B_f^{(1)}$ for $\mathbb{Z}_{N-4}$. This is accomplished by first viewing $B_c^{(1)}$ and $B_f^{(1)}$ as $U(1)$ gauge fields and writing
\be \tilde{a} = a + \onov{N} B_c^{(1)} \ \ \ {\rm and} \ \ \ \tilde{A}_f = A_f + \onov{N-4} B_f^{(1)} \nn\ee
We then further relate these to 2-form gauge fields in the usual manner for discrete gauge symmetries,
\be N B_c^{(2)} = dB_c^{(1)} \ \ \ {\rm and} \ \ \ (N-4) B_f^{(2)} = dB_f^{(1)} \nn\ee
Finally, we also introduce the $U(1)$ gauge field $A_2^{(1)}$ which will ultimately serve to gauge $(\mathbb{Z}_2)_F$. The symmetry structure \eqn{geven} is then implemented by various 1-form gauge transformations under which we have
\be \begin{aligned} B_c^{(2)} &\to B_c^{(2)} + d\lambda_c^{(1)} \\ B_c^{(1)} &\to B_c^{(1)} + N \lambda_c^{(1)} \end{aligned} \qquad \text{and} \qquad \begin{aligned} B_f^{(2)} &\to B_f^{(2)} + d\lambda_f^{(1)} \\ B_f^{(1)} &\to B_f^{(1)} + (N - 4) \lambda_f^{(1)} \end{aligned} \label{bshift}\ee
From this, it follows that
\be
\tilde{a} &\to \tilde{a} + \lambda_c^{(1)} \nn\\
\tilde{A}_f &\to \tilde{A}_f + \lambda_f^{(1)} \nn\\
A &\to A + \lambda_c^{(1)} + \lambda_f^{(1)} \nn\\
A_2^{(1)} &\to A_2^{(1)} + \frac{N}{2} \lambda_c^{(1)} + \frac{N-4}{2} \lambda_f^{(1)} \label{ashift}
\ee
By virtue of this last expression, the combination $A_2^{(1)}-B_c^{(1)}/2 -B_f^{(1)}/2$ is invariant under 1-form transformations and carries the holonomy appropriate for a $(\mathbb{Z}_2)_F$ gauge field by setting
\be 2A_2^{(1)} - B_c^{(1)} - B_f^{(1)} = dA_2^{(0)} \label{flat}\ee
for some periodic scalar $A_2^{(0)} \in [0, 2\pi)$. All of the equations above appear in \cite{bk2,bk1}.

Taking the exterior derivative of this equation, and dividing by 2, yields a local equation of 2-forms
\be dA_2^{(1)} - \frac{N}{2} B_c^{(2)} - \frac{N-4}{2} B_f^{(2)} = 0 \label{stillflat}\ee
To continue, we integrate \eqn{stillflat} over any 2-cycle $\Sigma$. This gives
\be \frac{N}{2} \int B_c^{(2)} + \frac{N-4}{2} \int B_f^{(2)} \ \in \ 2\pi \mathbb{Z} \label{constraintagain}\ee
where we have made use of the normalisation $\int dA_2^{(1)} \in 2\pi \mathbb{Z}$. The factor of $\frac12$ on the left-hand-side means that this equation should be viewed as a constraint, restricting the fluxes of the two discrete gauge fields. This mirrors the analysis of redundant symmetries and, in particular, the relation \eqn{alpha3} between $\mathbb{Z}_N$ and $\mathbb{Z}_{N-4}$ transformations where the same factor of 2 can be seen.

Suppose that we choose to set $B_f^{(2)} = 0$. The expression above then tells us that the normalisation of $B_c^{(2)}$, when integrated over a 4-cycle, is
\be \frac{N^2}{4} \int \frac{1}{2} (B_c^{(2)})^2 \ \in \ (2\pi)^2 \mathbb{Z} \ \ \ \Longrightarrow \ \ \ \frac{N^2}{8\pi^2}\int (B_c^{(2)})^2 \ \in \ 4 \mathbb{Z} \label{notodd}\ee
where the additional factor of $\frac{1}{2}$ on the left-hand side is simply a combinatoric factor.

\subsubsection*{The Putative Anomaly}

The discussion above differs in two places from that in \cite{bk1,bk2}. The first of these is not particularly consequential: the authors of \cite{bk1,bk2} insist that $dA_2^{(0)}$ is not closed because of the periodic nature of $A_2^{(0)}$. This is incorrect: $dA_2^{(0)}$ is closed, though it is not exact. This means that they have another term on the right-hand side of \eqn{stillflat}.

However, the key difference between the analysis above and that of \cite{bk1,bk2} lies in the normalisation of the fluxes \eqn{constraintagain}: their fluxes are normalised to $\int dA_2^{(1)} \in \pi \mathbb{Z}$ and, correspondingly, does not include the factor of $\frac12$ on the left-hand-side. (See, for example, equation (5.25) in \cite{bk2} for the symmetric theory, or (1.9) of \cite{bk1} for the anti-symmetric theory.) This incorrect choice of normalisation appears to be imposed because $A_2^{(1)}$ is a $\mathbb{Z}_2$ gauge field, but this conflates holonomy with curvature. It is this factor of 2 that resulted in the erroneous conclusion of a mixed anomaly.

To see how this incorrect normalisation gives the purported anomaly, it is simplest to set $B_f^{(2)}=0$ so the theory is coupled only to background gauge fields $A$, $A_2^{(1)}$ and $B_c^{(2)}$, associated to a nontrivial $\left(U(1)\times (\mathbb{Z}_2)_F\right)\big/\mathbb{Z}_N$ bundle. In the UV, there are 't Hooft anomalies due to the massless chiral fermions which can be summarised by a 5d WZW action containing the term
\be S_{WZW} = \frac{1}{8\pi^2}\int_{Y_5} N^2(B_c^{(2)})^2 A_2^{(1)},\nn\ee
where $Y_5$ is a 5-dimensional bulk whose boundary $\partial Y_5$ is the closed 4-dimensional manifold $X_4$ where our theory lives. Under a $(\mathbb{Z}_2)_F$ gauge transformation $\delta \deltafix A_2^{(1)} = \frac{1}{2} d \delta \deltafix A_2^{(0)}$, with $\delta \deltafix A_1^{(0)}=2\pi$, the anomaly action above changes by
\be \delta S_{WZW} = \frac{1}{8\pi^2}\int_{X_4} N^2 (B_c^{(2)})^2 \frac{\delta \deltafix A_2^{(0)}}{2} = \frac{N^2}{8\pi} \int_{X_4} (B_c^{(2)})^2\ee
Therefore, the partition function acquires a phase
\be \exp(i \delta S_{WZW}) = \exp\left(\pi i \frac{N^2}{8\pi^2} \int_{X_4} (B_c^{(2)})^2\right)\ee
which implies that the partition function flips sign when $(N^2/8\pi^2)\int (B_c^{(2)})^2$ is odd.

However, as seen in \eqn{notodd}, the correct normalisation does not permit $(N^2/8\pi^2)\int (B_c^{(2)})^2$ to be odd. Another way of saying this, as highlighted in both \eqn{alpha3} and \eqn{constraintagain}, is that $B_c^{(2)}$ and $B_f^{(2)}$ are not independent: we can have $(N^2/8\pi^2) \int (B_c^{(2)})^2$ odd, but only when $B_f^{(2)}\neq 0$. This correct normalisation nullifies the putative anomaly claimed in \cite{bk1,bk2}.

The lack of mixed anomaly in this theory can be understood in a number of different, complementary ways. In the remainder of this section we show how one can reach the same conclusion from different viewpoints.

\subsection{Line and Surface Operators}

A useful, physical rephrasing of the story above can be seen in the language of topological line and surface operators in the spirit of \cite{sk}. First, consider the field $A_2^{(1)}$. It is subject to the following gauge transformations
\be A_2^{(1)} \to A_2^{(1)} + \frac{N}{2} \lambda_c^{(1)} + \frac{N-4}{2} \lambda_f^{(1)} + d\lambda_2^{(0)} \nn\ee
where $\lambda_2^{(0)} \in [0, 2\pi)$ is a compact scalar.
This states that although the line operator $\exp(i\int A_2^{(1)})$ isn't gauge-invariant on its own, $N/2$ surface operators $\exp(i\int B_c^{(2)})$ and $(N-4)/2$ surface operators $\exp (i\int B_f^{(2)})$ can end on it to make it gauge-invariant. Furthermore, by virtue of the constraint \eqn{stillflat}, this junction is topological; the line can be moved around to eat up the surfaces. It follows that the surface operator
\be \exp\left( i \int B_c^{(2)} \right)^{N/2 } \, \exp\left( i \int B_f^{(2)} \right)^{(N-4)/2} \label{trivsurface}\ee
is trivial, for we can open up a hole with the line operator $\exp(i\int A_2^{(1)})$ on the boundary, and continue enlarging this hole until the surfaces disappear.

The triviality of \eqn{trivsurface} leads to the same constraint on $B_c^{(2)}$ and $B_f^{(2)}$ that we saw earlier. Note however this argument highlights the fact there was no subtlety hiding in the division by 2 that took us from \eqn{flat} to \eqn{stillflat}. Instead, the crucial fact was simply the ability to write down the line operator $\exp(i\int A_2^{(1)})$.

\subsection{Cocycle Conditions}

It is possible to analyse the symmetry structure \eqn{geven} using discrete gauge fields. This has the advantage of showing in a very simple way how the condition \eqn{constraintagain} arises, as well as being more natural from a mathematical standpoint. In the discrete formulation, the $\mathbb{Z}_2$ gauge field is described by the transition functions
\be s_{ij} \in \mathbb{Z}_2 \nn\ee
In normal circumstances, these transition functions would obey the cocycle condition $s_{ij} + s_{jk} + s_{ki} = 0 \text{ mod } 2$. But in the presence of the 2-form gauge fields $B_c^{(2)}$ and $B_f^{(2)}$, the cocycle condition is instead relaxed to
\be s_{ij} + s_{jk} + s_{ki} = B^c_{ijk} + B^f_{ijk} \mod 2 \nn\ee
Here $B^c_{ijk} \in \mathbb{Z}_N$ and $B^f_{ijk} \in \mathbb{Z}_{N-4}$ are the cocycles describing the 2-form gauge fields. The above equation, when reduced to cohomology, defines an equation in $H^2(M; \Z_2)$. Moreover, the left hand side is the definition of a coboundary, so becomes trivial in cohomology. Therefore the above equation becomes
\be 0 = B^c + B^f \mod 2 \nn\ee
The relation between discrete and continuum gauge fields is, roughly speaking, $\frac{2\pi}{N} B^c = B_c^{(2)}$ and $\frac{2\pi}{N - 4} B^f = B_f^{(2)}$. Recasting the previous equation into continuum normalisation, it becomes
\be 0 = \frac{N}{2} B_c^{(2)} + \frac{N-4}{2} B_f^{(2)} \mod 2\pi \nn\ee
This is the promised discrete analogue of \eqn{constraintagain}.

\subsection{Cobordism}\label{boredsec}

So far we have considered anomalies in the internal symmetry group $G$, but made no mention of metrics or gravitational anomalies. The right framework to discuss these anomalies is to consider a theory on a curved manifold with a Spin-$G$ structure. Furthermore, the right way to think about anomalies for discrete symmetries is in terms of cobordism. If there is an anomaly it should also show up in the cobordism approach where we couple to a Spin-$G$ structure. We will now see that no such anomaly exists.

We again restrict attention to the theory with anti-symmetric matter, although similar remarks apply to the symmetric theory. The group $\text{Spin}_G$ is defined by
\be \text{Spin}_G = \frac{G \times \text{Spin}(4)}{(\mathbb{Z}_2)_F} = \frac{SU(N - 4) \times U(1) \times \text{Spin}(4)}{\mathbb{Z}_N \times \mathbb{Z}_{N-4}} \nn\ee
The putative anomaly between $(\mathbb{Z}_2)_F$ and $\mathbb{Z}_N$ can be seen even when the $SU(N - 4)$ gauge field is set to zero. Therefore it should be possible to see the anomaly working only with the subgroup
\be \frac{U(1) \times \text{Spin}(4)}{\mathbb{Z}_N} \subset \text{Spin}_G \nn\ee
The latter is isomorphic to
\be \frac{U(1) \times \text{Spin}(4)}{\mathbb{Z}_2} = \text{Spin}_c \nn\ee
via the explicit isomorphism $(e^{i \theta}, S) \rightarrow (e^{i (N/2) \theta}, S)$. This means the anomaly should be visible in the coupling of the theory to a background $\text{Spin}_c$ structure. However, since
\be \text{Tor} \, \Omega_5^{\text{Spin}_c}(\text{pt}) = 1 \nn\ee
is trivial \cite{garcia}, $\text{Spin}_c$ can only have perturbative anomalies, not global ones. Since the UV and IR have matching perturbative anomalies, we conclude they are fully consistent.

\section{A Field Theoretic Perspective} \label{otherargs}

The previous arguments are rather mathematical in nature. Here we present a more physical, field theoretical approach to the problem. The question we wish to address is whether the low-energy phases with massless fermions and unbroken global symmetry $G$ share the same anomalies as the UV theory. One foolproof way to demonstrate this is to exhibit a path, preserving the symmetry, from the UV theory to the free fermion phase. In this section we exhibit such paths, first for the anti-symmetric theory and then for the symmetric theory.

\subsection{The Anti-Symmetric Theory}\label{antisec}

In this section we describe a (mostly) weakly coupled phase of the anti-symmetric theory in which the low-energy physics consists of a massless fermion with unbroken symmetry group \eqn{geven}. This leaves the would-be anomaly nowhere left to hide. The arguments below closely follow those of \cite{raby,seibtalk}, with the only novelty a check that the discrete quotients are unaffected by these arguments.

We start by adding some gauge and flavour indices to our fermions: these are $\chi_{ij}$ and $\psi^i_{\ a}$ where $i=1,...,N$ is the gauge index and $a=1,...,N-4$ the flavour index under the $SU(N-4)$ global symmetry.

We next introduce $N-4$ fundamental scalars $\phi^{ia}$. Clearly there is no obstacle to giving these a mass and they do not affect the anomalies. If the fermions $\psi$ are taken to transform in the $(\overline{\Box}, \Box)$ of $SU(N) \times SU(N-4)$, then we take the scalars to sit in the $(\overline{\Box}, \overline{\Box})$ representation. We assign them $U(1)$ charge $q_\phi = 1$. The following Yukawa interaction is then invariant under all symmetries,
\be \lag_{Y}\sim \phi\psi\chi+c.c.\label{yuk}\ee
The gauge and global center transformations act on $\phi$ as
\be \phi\to e^{-2\pi im/N}e^{-2\pi ik/(N-4)}e^{2\pi i\al}\phi\nn\ee
One can check that the addition of the Yukawa terms does not affect the symmetry structure \eqn{geven}, including the discrete quotients.

We now give an expectation value to the scalars of the form
\be \vev{\phi^{ia}}=v\delta^{ia}\label{vev}\ee
Crucially, the full global symmetry survives, but is now locked with the gauge symmetry. To see this, note that the most general $SU(N)\times SU(N-4)\times U(1)$ transformation acting on $\phi$ can be written as
\be \phi\to e^{2\pi i\al}U^*\phi V^\dagger\ ,\ U\in SU(N)\ ,\ V\in SU(N-4)\nn\ee
The choice
\be U=\mat{V^* e^{2\pi i\al}&\\&U'e^{-\frac{2\pi(N-4)i\al}{4}}}\ ,\ U'\in SU(4)\nn\ee
leaves the vacuum invariant. Because the global symmetry survives, so too do its 't Hooft anomalies and these can be matched between the UV and IR. Meanwhile, the gauge symmetry is Higgsed down to $SU(4)$. We will look more closely at the fate of this $SU(4)$ below.

In the Higgs phase, all the scalars are gapped. The Yukawa interaction \eqn{yuk} gives a mass term of the form $v\sum_{a=1}^{N-4}\sum_{i=1}^N\chi^{ai}\psi^{ai}$. This gives a mass to most of the fermions. The only ones that survive transform under $SU(4) \times SU(N-4)\times U(1)$ as
\be
\mbox{$\Psi$ in $({\bf 1},\sym)_{-N/2}$:} &\quad \mbox{This comes from the fermion $\psi$} \nn\\
\mbox{$\rho$ in $({\bf 6},{\bf 1})_0$:} &\quad \mbox{This comes from the fermion $\chi$} \nn
\ee
We now want to consider more carefully how the various discrete gauge fields couple. To this end, let us briefly return to the un-Higgsed UV theory and couple it to background gauge fields for the symmetry $G_{\rm even}$ defined in \eqn{geven}. It will be useful to write explicitly all the covariant derivatives of the fields including the background fields. Denote by $\tilde{a}=a+\onov{N}B_c^{(1)}$ the promoted $SU(N)\to U(N)$ gauge field, and $\tilde{A}_f+\onov{N-4}B_f^{(1)}$ the promoted $SU(N-4)\to U(N-4)$ gauge field. Meanwhile, the $U(1)$ gauge field is written as $A$, and the $(\mathbb{Z}_2)_F$ as $A_2^{(1)}$. Under 1-form gauge transformations, we have
\be \tilde{a}&\to\tilde{a}+\la_c\nn\\ \tilde{A}_f&\to\tilde{A}_f+\la_f\nn\\ A&\to A+\la_c+\la_f\nn\\ A_2&\to A_2+\frac{N}{2}\la_c+\frac{N-4}{2}\la_f\nn\ee
One can check that the corresponding covariant derivatives are:
\be &D\psi=(\partial-iR_{\bar{F}}(\tilde{a})-iR_F(\tilde{A}_f)+i(N-2)/2A-iA_2^{(1)})\psi\nn\\
&D\chi=(\partial-iR_A(\tilde{a})-i(N-4)/2A+iA_2^{(1)})\chi\nn\\
&D\phi=(\partial-iR_{\bar{F}}(\tilde{a})-iR_{\bar{F}}(\tilde{A}_f)-iA)\phi\ \nn
\ee
where $R_X(a)$ the gauge field in representation $X$. 
This is before the Higgsing. After the scalar gets an expectation value \eqn{vev}, we write the $SU(N)$ gauge field as
\be a=\mat{-A_f+\left(A-\onov{N}B_c^{(1)}-\onov{N-4}B_f^{(1)}\right)\mathbf{1}_{N-4}&\\&a_4-\frac{N-4}{4}\left(A-\onov{N}B_c^{(1)}-\onov{N-4}B_f^{(1)}\right)}\nn\ee
where $a_4$ is the $SU(4)$ gauge field that survives the Higgs mechanism, and we set all massive fields to zero. This equation fixes $a$ such that the vacuum is invariant under the global symmetries. In other words, it gives us the colour-flavour locking pattern. The covariant derivatives of the surviving massless fermions are
\be &D\Psi=\left(\partial-iR_S(\tilde{A}_f)+\frac{iN}{2}A-iA_2^{(1)}\right)\Psi\nn\\ &D\rho=\left(\partial-iR_A(a_4)-\frac{i}{2}B_c^{(1)}-\frac{i}{2}B_f^{(1)}+iA_2^{(1)}\right)\rho\nn\ee
Up to this point, everything is weakly coupled. There is no place for the anomalies to hide: because the global symmetry survives the Higgs mechanism unscathed, with only a gauge twist for its troubles, all 't Hooft anomalies of the UV theory must be reproduced in the IR by the massless fermions $\Psi$ and $\rho$.

At first glance, the appearance of $B_c^{(1)},\ B_f^{(1)},\ A_2^{(1)}$ in the covariant derivative of $\rho$ may suggest something that looks like an anomaly related to these three fields. However, $\rho$ cannot carry any anomaly. This follows simply because it is gappable: it sits in a real representation of $SU(4)$ and a mass term $\rho\rho$ breaks no symmetry. Furthermore, $(-1)^F$ acting on $\rho$ is nothing but an $SU(4)$ gauge transformation therefore we know that there cannot be an associated anomaly. The upshot is that both the $SU(4)$ sector, with the fermion $\rho$ become massive and decouple from the infra-red physics.

We are left, in the infra-red, with the massless fermion $\Psi$. This carries the same quantum numbers as the composite fermion $\chi\psi\psi$ and so the anomalies in the two phases necessarily agree: both are described by a massless fermion in the $(\sym)_{-N/2}$ under $SU(N-4)\times U(1)$. This argument, which is largely a recapitulation of standard lore, is the reason why the confining and Higgs phases are thought to coincide in the anti-symmetric theory\footnote{An attempt to distinguish the two phases was also made in \cite{bkl3} through the study of the anomalous global $U(1)$ symmetry, with the argument that in the confining phase there is no would-be scalar operator that is charged only under this (non)-symmetry. We do not comment on the validity of the argument, but simply mention that the baryonic state $B=\chi^{N-2}\psi^{N-4}$ does the job. Its colour, flavour and Lorentz indices are all contracted to give a scalar. Its condensate $\vev{B}\neq 0$ doesn't break any symmetry, but it is ``charged" under the anomalous $U(1)$.}.

\subsection{The Symmetric Theory}

We now turn to the symmetric theory. In this case, the confining and Higgs phases clearly differ. While the massless fermion $\lambda\psi\psi$ saturates the perturbative 't Hooft anomalies of the $SU(N+4) \times U(1)$ global symmetry, one may wonder if it also saturates discrete anomalies, or perhaps even anomalies that are still to be discovered. Our goal here is to exhibit a path in field space that makes it clear that the phase with unbroken global symmetry and a massless fermion is indeed consistent.

To achieve this, we start with a different UV theory. We will show that as we dial a parameter, we can interpolate between the symmetric chiral gauge theory, and the free massless fermion.

The parent theory has gauge group $H=SU(N)\times SU(N-4)$, with left-handed Weyl fermions in the following representations:
\be
\mbox{$\chi$ in $\left(\anti,\bf 1\right)$ $\ ,\quad$}
\mbox{$\psi$ in $(\overline{\Box},\Box)$ $\ ,\quad$}
\mbox{$N$ copies of $\eta$ in $(\bf 1,\overline{\Box})$}
\ee
The theory has a global symmetry
\be G =SU(N) \times U(1) \nn\ee
Here only the $\eta$ transform under the global $SU(N)$, while the $U(1)$ charges are $q_\chi=N-4$ and $q_\psi=-q_\eta=2-N$. (As before, this $U(1)$ acts faithfully only when $N$ is odd; for $N$ even, the charges should be divided by 2.)

Associated to each factor in the gauge group $H$ is a strong coupling scale, $\Lambda_N$ and $\Lambda_{N-4}$, with the ratio
\be x = \frac{\Lambda_N}{\Lambda_{N-4}}\nn\ee
We can analyse the theory in the two limits $x\rightarrow 0$ and $x\rightarrow \infty$. We will show that one limit reproduces the symmetric chiral gauge theory, while the other results in the free massless fermion. Note that we make no claims about the dynamics as one varies $x$ and it may well be that a phase transition separates the two ends. Our purpose here is simply to make a much weaker kinematical statement: the massless fermion of the confined phase is a consistent possibility for the low-energy physics of the symmetric theory.

\noindent
\underline{$x\rightarrow \infty$:}

\noindent
In the limit $x\rightarrow \infty$, the $SU(N)$ gauge group is the first to become strongly coupled. This is a copy of the anti-symmetric theory that we have just discussed. Of course, we do not know the low-energy physics but that is not our immediate concern: all that we care about is that the confining phase of this theory, with its associated massless fermion $\Psi =\chi\psi\psi$, is consistent. (Indeed, we could add further Higgs fields as in Section \ref{antisec} to ensure that we sit in this phase.)

At energy scales $\Lambda_{N-4}\ll E\ll\Lambda_{N}$, we are left with an $SU(N-4)$ gauge theory coupled to the composite fermion $\Psi$, transforming in the $\sym$ and $N$ fermions $\eta$ in the $\overline{\Box}$. This, of course, is the symmetric chiral gauge theory.

\newpage
\noindent
\underline{$x\rightarrow 0$:}

\noindent
In the opposite limit, $x\to 0$, the $SU(N-4)$ theory first becomes strong. The dynamics of this theory is QCD-like, with $N$ Dirac fermions and the expectation is that the fermion bilinear $\psi\eta$ condenses, resulting in an $SU(N)$ non-linear sigma model at low-energy, parametrised by the field $U\in SU(N)$.

At energy scales $\Lambda_N\ll E\ll\Lambda_{N-4}$, we are left with an $H'=SU(N)$ gauge theory coupled to a single fermion $\chi$ in the $\anti$ and the Goldstone mode $U$. Importantly, this Goldstone mode is charged under the $H'=SU(N)$ gauge symmetry as $U\to U'=VU$ with $V\in H'$. As a result, the vacuum of the sigma model acts to completely Higgs the $H'=SU(N)$ gauge group. The $G=SU(N)\times U(1) $ global symmetry is preserved, albeit only after mixing with the $H'=SU(N)$ gauge symmetry. The upshot is that both the gauge field and the Goldstone mode $U$ become gapped, leaving us with the massless fermion $\chi$ and an unbroken global symmetry. This is the low-energy physics of the confined phase of the symmetric chiral gauge theory.

Again, we stress that we make no claims that the confined phase is dynamically realised by the symmetric chiral gauge theory: only that 't Hooft anomalies, perturbative or otherwise, present no obstacle to doing so.

\acknowledgments
We thank Stefano Bolognesi, Ken Konishi and Andrea Luzio for communications. We are also grateful to Pietro Benetti Genolini, Joe Davighi, Kaan Onder, Carl Turner and Shimon Yankielowicz for many discussions on these issues. AK would also like to thank the Blavatnik family foundation for the generous support. We are supported by the STFC consolidated grant ST/P000681/1 and a Simons investigator award. AK is supported by the Blavatnik postdoctoral fellowship and DT by a Wolfson Royal Society Research Merit Award.

\end{document}